\documentclass[twocolumn,tighten,times]{aastex63}
\usepackage{amsmath}
\usepackage[caption=false]{subfig}

\newcommand{\Ms}{sMBH}
\newcommand{\Mp}{pMBH}

\newcommand{\Msb}{$M_{\mathrm{sb}}$}

\newcommand{\D}{${\cal D}$}

\newcommand{\q}{$q$}

\newcommand{\Rbh}{$R_{\mathrm{b,h}}$}
\newcommand{\Rgh}{$R_{\mathrm{g,h}}$}

\newcommand{\lisa}{\textit{LISA}}
\newcommand{\zdagn}{$z_{\mathrm{dAGN}}$}
\newcommand{\zcoal}{$z_{\mathrm{coal}}$}

\submitjournal{ApJ}

\shorttitle{Using Dual AGNs to Predict the Rate of Massive Black Hole Mergers}
\shortauthors{Li et al.}

\begin{document}

\title{Massive Black Hole Binaries from the TNG50-3 Simulation: II. Using Dual AGNs to Predict the Rate of Black Hole Mergers} 
%Investigating the Roles of Dynamical Friction and Radiation Feedback

\author[0000-0002-0867-8946]{Kunyang Li}
\affiliation{School of Physics and Center for Relativistic
  Astrophysics, 837 State St NW, Georgia Institute of Technology,
  Atlanta, GA 30332, USA}   
\email{kli356@gatech.edu}

\author[0000-0002-7835-7814]{Tamara Bogdanovi{\'c}}
\affiliation{School of Physics and Center for Relativistic
  Astrophysics, 837 State St NW, Georgia Institute of Technology,
  Atlanta, GA 30332, USA}
\email{tamarab@gatech.edu}

\author[0000-0001-8128-6976]{David R. Ballantyne}
\affiliation{School of Physics and Center for Relativistic
  Astrophysics, 837 State St NW, Georgia Institute of Technology,
  Atlanta, GA 30332, USA}
\email{david.ballantyne@physics.gatech.edu}

\author[0000-0001-8128-6976]{Matteo Bonetti}
\affiliation{Dipartimento di Fisica G. Occhialini, Università di
  Milano-Bicocca, Piazza della Scienza 3, IT-20126 Milano, Italy}
\affiliation{INFN, Sezione di Milano-Bicocca, Piazza della Scienza 3, IT-20126 Milano, Italy}
\email{matteo.bonetti@unimib.it}

%\correspondingauthor{Kunyang Li}
%\email{kli356@gatech.edu}

\begin{abstract}
Dual active galaxy nuclei (dAGNs) trace the population of post-merger
galaxies and are the precursors to massive black hole (MBH) mergers,
an important source of gravitational waves that may be observed by
\lisa. In Paper I of this series, we used the population of $\approx 2000$~galaxy
mergers predicted by the TNG50-3 simulation to seed semi-analytic
models of the orbital evolution and coalescence of MBH pairs with
initial separations of $\approx 1$~kpc. Here, we
calculate the dAGN luminosities and separation of these pairs as they evolve in post-merger galaxies, and show how the
coalescence fraction of dAGNs changes with redshift. We find that
because of the several Gyr long dynamical friction timescale for orbital
evolution, the fraction of dAGNs that eventually end in a MBH merger
grows with redshift and does not pass 50\% until \zdagn$\approx
1$. However, dAGNs in galaxies with
bulge masses $\la 10^{10}$~M$_{\odot}$, or comprised of near-equal
mass MBHs, evolve more quickly and have higher than average
coalescence fractions. At any redshift, dAGNs observed with small
separations ($\la 0.7$~kpc) have a higher probability of merging in a
Hubble time than more widely separated systems. As found in Paper I,
radiation feedback effects can significantly reduce the number of MBH
mergers, and this could be manifested as a larger than expected number of
widely separated dAGNs. We present
a method to estimate the MBH coalescence rate as well as the potential
\lisa\ detection rate given a survey of dAGNs. Comparing these rates
to the eventual \lisa\ measurements will help determine the efficiency of
dynamical friction in post-merger galaxies.

\end{abstract}


%%%%%%%%%%%%%%%%%%%%%%%%%%%%%%%%%%%%%%\
%  SECTION 1
%%%%%%%%%%%%%%%%%%%%%%%%%%%%%%%%%%%%%%\
\section{Introduction}
\label{sec:intro}
Dual Active Galactic Nuclei (dAGNs) are two accreting massive black
holes (MBHs) residing within a single host galaxy and are expected to
occur following the merger of two massive galaxies. A
population of dAGNs in post-merger galaxies seems to be an unavoidable prediction of hierarchal galaxy
formation models \citep[e.g.][]{Rosa2020}. In some cases, the separation of the two
MBHs that comprise a dAGN will shrink slowly over time as the orbiting
MBHs interact with the gaseous and stellar backgrounds of the galaxy
\citep[e.g.,][]{BBR1980,Volonteri2003, B2012, V2016,Bonetti2019,
  K2019, LBB22a}, eventually leading to the emission of gravitational
waves and the coalescence of the two black holes
\citep[e.g.,][]{LISA2017,KBH2017,Kelley2019}. Therefore,
dAGNs are `tracers' of future MBH mergers and gravitational wave
events. An observed sample of dAGNs, combined with a
model describing their future evolution, can thus provide a prediction of the MBH merger rate, a critical parameter for the
upcoming \textit{Laser Interferometer Space Antenna} (\lisa) gravitational wave
observatory. 

%Massive black holes (MBHs) are known to reside in the centers of most massive galaxies \citep{S1982, KR1995,M1998}, and the hierarchal model of galaxy evolution predicts that massive galaxies are built up through series of mergers \citep[e.g.,][]{White1978,White1991}. Most mergers generate significant nuclear gas
%flows \citep[e.g.,][]{D2005} that provide a favorable environment in which the
%MBHs can accrete and shine as AGNs. Therefore, a population of dual
%AGNs (dAGNs) in post-merger galaxies seems to be an unavoidable prediction of
%hierarchal galaxy formation \citep[e.g.,][]{Rosa2020}. In addition, dAGNs are expected to be the parent population of binary MBHs, where the two MBHs are gravitationally
%bound and the orbit decays through the emission of gravitational waves \citep[e.g.,][]{BBR1980,LISA2017,KBH2017,Kelley2019}. Through interactions with the stellar and gaseous background of the remnant galaxy, some MBH pairs sink toward the galactic core,  coalesce, and become powerful gravitational waves (GWs) sources. They are key targets of the Laser Interferometer Space Antenna (\lisa) which will survey the frequency range of $100 \mu {\rm Hz} \sim 100 {\rm mHz}$ \citep{K2016, LISA2017, KL2018}. Thus, understanding the population of dAGNs is necessary for determining the expectations for
%future gravitational wave experiments.

Whether an observable dAGN results in a MBH coalescence depends on the physical
processes within the remnant galaxy that drive the orbital evolution
of the MBHs. When the MBHs are at separations of $\sim 1$~kpc,
dynamical friction (DF) by gas and stars is expected to dominate the
orbital decay \citep{BBR1980}. In this process, gravitational
deflection of gas \citep{O1999,KK2007} or collisionless particles
\citep[e.g., stars and dark matter;][]{C1943,AM2012} causes an
overdense wake to form behind each MBH. The wakes exert a gravitational pull onto the MBHs, which saps their orbital energy. Once
the two MBHs are gravitationally bound (at separations $\sim$~pc)
stellar ``loss-cone"  scattering is expected to dominate the orbital
decay \citep[e.g.,][]{Q1996, QH1997, Y2002}. If the
galaxy is sufficiently gas rich, drag on the binary by the surrounding
circumbinary disk may also affect its orbital evolution at separations
$\la 0.1$~pc \citep[e.g.,][]{A2005, MP2005}. Only when the separation
falls below $\sim 1000$ Schwarzschild radii does gravitational wave
emission begin to dominate the orbital evolution until the MBHs merge
\citep[e.g.,][]{KT1976, BBR1980}. In addition to the the orbital decay, the gas content of the post-merger galaxy will also
  strongly influence the accretion rates onto the MBHs and their subsequent
  properties as dAGNs. As a result, connecting observations of dAGNs to their
  potential future gravitational wave sources will depend on the
  properties of the host galaxy and the orbit of the MBH pair
  \citep[e.g.,][]{LBB21a}. 

In Paper I of this series \citep{LBB22a}, we presented the results of
a semi-analytic model that followed the dynamical evolution of
$\approx 8000$ MBH pairs from $\sim 1$~kpc to coalescence accounting for all the
processes described above (DF, loss-cone scattering,
decay in a circumbinary disk, and gravitational wave emission). The host galaxy models in
which the MBH pairs evolved were constructed from the properties of
merger galaxies in the TNG50-3 cosmological
simulation \citep{Nelson2019, TNG50_a, TNG50_b}. Paper I showed that
the DF phase was the most important process in determining if a MBH
pair would coalesce within a Hubble time, and therefore the stellar
and gas content at scales of a few hundreds pc in post-merger galaxies are critical
for the expected \lisa\ detection rates. Moreover, we found that
radiation feedback effects from the accreting MBHs has the potential
to significantly weaken DF forces in gas-rich galaxies
\citep[e.g.,][]{PB2017,LBB20b}, which could severely reduce the number
of MBH mergers detected by \lisa. It is crucial, therefore, to find a
method that can test these predictions, especially the potential role
of radiation feedback, and observationally constrain the
\lisa\ expectations.

Here, we build on the results of Paper I by describing the expected
dAGN properties of MBH pairs as they evolve towards
coalescence. Paper I shows that the galaxy properties at $\sim
1$~kpc scale are often crucial for the overall orbital decay, and so we focus
on the dAGN properties with separations in this range. As our model
follows these MBH pairs through to coalescence, we demonstrate that
observational surveys of dAGNs can provide an estimate of the expected
MBH merger rate. Crucially, we consider the impact of radiation
feedback on both the dynamics and accretion onto the MBHs, allowing
for an observational determination of the magnitude of this effect.

This paper is organized as follows. In \S~\ref{sec:methods} we provide
a brief summary of the main features of the calculation used to evolve the MBHs 
and how the TNG50-3 simulation data is used as input to the
model. \S~\ref{sec:AGN_dist_zla} shows how the distributions of
dAGN luminosity and separation change over time.
\S~\ref{sec:AGN_coal} presents the merger fraction of dAGNs at
different redshifts and how this depends on the properties of the host
galaxy and MBH pair.
%In \S~\ref{sec:Detect} we discuss the detectability of dAGNs via
%direct imaging.
The impact of radiation
feedback on these results is shown in \S~\ref{sec:RF_AGN_dist}. Finally, we discuss the
implications of our findings in \S~\ref{sec:discuss} and conclude in
\S~\ref{sec:concl}. We assume a cosmology consistent with that used in the TNG50-3 simulation ($\Omega_{\rm \Lambda,\,0}=0.6911$, $\Omega_{\rm m,\, 0}= 0.3089$, $\Omega_{\rm b, \, 0}= 0.0486$, $h=0.6774$), and $t_{\rm Hubble}=14.4$ billion yrs.

%%%%%%%%%%%%%%%%%%%%%%%%%%%%%%%%%%%%%%\
%  SECTION 2
%%%%%%%%%%%%%%%%%%%%%%%%%%%%%%%%%%%%%%\
\section{Methods}
\label{sec:methods}
A thorough description of the dynamical evolution calculation of MBH
pairs and our use of the TNG50-3 simulation data is found in Paper I (see also
\citealt{LBB20a}). We therefore provide a brief summary of the method
below before describing how we compute the time-dependent accretion
rate and luminosity of each of our model dAGNs.

%We run our model on the close MBH pairs drawn from the cosmological simulation TNG50 \citep{Nelson2019, TNG50_a, TNG50_b}, the third and final installment of the IllustrisTNG project \citep{Naiman2018, Nelson2018, Marinacci2018, Pill2018,Spring2018}. In this section, we first give a brief overview of the model of the merger remnant galaxy and then briefly describe the TNG50 data used as the input for our semi-analytic model.

%%%%%%%%%%%%%%%%%%%%%%%%%%%%%%%%%%%%%%\
%  SECTION 2.1
%%%%%%%%%%%%%%%%%%%%%%%%%%%%%%%%%%%%%%\
\subsection{The Dynamical Evolution of MBH Pairs in TNG50-3
  Post-Merger Galaxies}
\label{sec:galaxymodel}
We assume that a galaxy merger produces a single remnant, with a
stellar bulge and gas disk\footnote{We neglect the stellar disk in the
calculation as its impact on the orbital evolution of an MBH is
relatively minor \citep{LBB20a}.}, which includes the MBH pair. The primary MBH (\Mp; with mass $M_{\rm 1}$) is fixed at the center of the galaxy. The non-rotating bulge has a mass $M_{\rm sb}$ and
follows a coreless powerlaw density profile \citep[e.g.,][]{BT1987}, which is cutoff at twice the half-mass radius of the bulge ($2\times$\Rbh), with the scale parameters proportional to $\log \,(M_{\rm 1}/10^{\rm 5} M_{\rm \odot})$ kpc. We consider the orbital
evolution of a bare, secondary MBH (\Ms; with mass $M_{\rm 2} < M_{\rm 1}$) which is orbiting in the plane of
the gas disk. The total mass of the MBH pair is $M_{\mathrm{bin}}=M_{\rm 1}+M_{\rm 2}$ and the mass ratio is $q=M_{\rm 2}/M_{\rm 1}$.

The gas fraction of the remnant
galaxy is $f_{\mathrm{g}}=M_{\rm gd}/(M_{\rm gd}+M_{\rm sb})$, where
$M_{\rm gd}$ is the mass of the gas disk within twice the half-mass
radius of the gas disk ($2\times$\Rgh). Once $f_{\mathrm{g}}$ is set
the gas densities are determined using an exponential profile with a scale
radius defined as $2 \log\,(M_{\rm 1}/10^{\rm 5}\, M_{\rm
  \odot})$~kpc \citep[e.g.,][]{BT1987}. As a result, galaxies with a
larger \Mp\ have gas densities that decrease more slowly with radius. The
gas disk of each galaxy rotates with a speed drawn from the uniform distribution $0.7-0.9 v_{\rm c}(r)$, where $v_{\rm c}(r)$ is the
local circular velocity.  

A list of 1997 galaxy merger events, including redshifts and MBH masses, is
extracted from the catalogs of the TNG50-3 simulation\footnote{See
\url{https://www.tng-project.org/data/docs/specifications
}}. Specifically, the redshifts correspond to when the two MBHs reach
a separation equal to the gravitational softening length of the
collisionless component ($\approx 1$~kpc). The properties of the
remnant galaxy ($M_{\rm sb}$, $M_{\rm gd}$, \Rbh, and \Rgh) are also
extracted from the TNG50-3 catalogs and are used to construct the
galaxy model within which the MBH pair evolves. The dynamical
evolution of the pair is initialized so that the semi-major axis $a$
is $\approx 1$~kpc. The initial eccentricity of the \Ms\ is set to be
either $e_i < 0.2$ or $0.8 \leq e_i \leq 0.9$, and we consider both
prograde and retrograde orbits \citep{LBB20a}. Thus, we compute four distinct
evolutions of the \Ms\ in each of the 1997 post-merger galaxies. The
results presented below are from the combined dataset of 7988 calculations.

The orbital evolution of the \Ms\ due to DF is computed as described
by \citet{LBB20a, LBB20b}. This process takes the \Ms\ down
to the influence radius of the MBH pair, where the mass enclosed by
the orbit is equal to twice the pair mass. Below this radius, the
orbital decay is due to the combination of loss-cone scattering, drag
from the circumbinary gas disk, and gravitational wave emission. The
calculation ends when the orbital separation is smaller than the
radius of the innermost stable circular orbit (ISCO) of a non-spinning
MBH pair (i.e., $R_{\rm ISCO} = 6 G M_{\rm bin}/
c^{2}$). Paper I provides full details of
how the orbital evolution is computed below the influence radius. The
full decay time of the \Ms\ is tracked and the MBH coalescence
redshift, \zcoal, is recorded for each calculation that merges within
a Hubble time ($36$\% of the 7988 orbital evolutions do not
successfully merge within this time; see \S~3 of Paper I).

Lastly, for those models which successfully reach coalescence, we
compute the expected \lisa\ signal-to-noise ratio (SNR) during the
inspiral phase assuming a four-year mission lifetime (Appendix A of
Paper I; \citealt{Bonetti2019}). The detection threshold
for \lisa\ used in this paper is a SNR$> 8$. This information will
allow us to connect \lisa-detectable MBH pairs to their earlier dAGN
properties.

\subsection{Accretion Rates and Luminosities}
\label{sub:accretion}
%The fixed \Mp\ and the moving \Ms\ are both allowed to accrete matter
%from their surroundings and thus, may appear as AGNs.
The accretion rates onto both the \Mp\ and \Ms\ are calculated as a
function of time during the DF phase of the calculation. That is, we
compute the dAGN luminosities only when the separations are greater
than the influence radius of the MBH pair ($\sim 1$~pc). For simplicity, we neglect
the mass increase of each MBH due to accretion and consider only
bolometric luminosities (see \S~\ref{sec:discuss}).

The accretion rate onto the
stationary \Mp\ is computed using the Bondi formula \citep{BH1944,
  Bondi1952} and its luminosity is limited to be no more than $10$\%
of the Eddington luminosity \citep[e.g.,][]{Lusso2012}, i.e.,
\begin{eqnarray}
  \label{eq:L1}
L_{1} =
\begin{cases}
  0.1 \dot{M}_{\rm B1} c^2 & {\rm when} \;\;\; L_{1} < 0.1L_{\rm 1, Edd}, \\
  0.1 L_{\rm 1, Edd} & \mathrm{otherwise},
\end{cases}
\end{eqnarray}
where $\dot{M}_{\rm B1} = \pi\, n_{\rm gd0}\, m_{\rm p}\,(GM_{\rm
  1})^2/c_{\rm s1,\infty}^3$ is the Bondi accretion rate onto the
\Mp\ and $L_{\rm 1, Edd} = 4\pi G M_{\rm 1}m_{\rm p}c / \sigma_{\rm
  T}$ is its Eddington luminosity. In the Bondi formula $n_{\rm gd0}$
is the central gas density and
$c_{\rm s1,\infty}$ is the sound speed at the galactic center. To
determine the sound speed, the temperature profile of the gas disk is
assumed to be $10^4$~K above the minimum Toomre stability temperature
\citep{T1964}. 

As the \Ms\ is moving through the post-merger galaxy, its accretion
rate is calculated using the Bondi-Hoyle-Lyttleton model, which accounts
for the drop in accretion due to the relative motion of the MBH \citep{HL1939, BH1944, Bondi1952},
$\dot{M}_{\rm BHL} = \dot{M}_{\rm B2} / (1+\Delta v^2/c_{\rm s2,
  \infty}^2)^{3/2}$. Here, $\dot{M}_{\rm B2}$ represents the regular
Bondi rate of the \Ms, $c_{\rm s2, \infty}$ is the sound speed of the
gas at the same radius as the \Ms, and $\Delta v$ is the velocity of
the \Ms\ relative to the gas disk. The resulting accretion luminosity
of the \Ms\ is 
\begin{eqnarray}
  \label{eq:L2}
L_{2} =
\begin{cases}
  0.1 \dot{M}_{\rm BHL} c^2 & {\rm when} \;\;\; L_{2} < L_{\rm 2, Edd}, \\
  L_{\rm 2, Edd} & \mathrm{otherwise},
\end{cases}
\end{eqnarray}
where $L_{\rm 2, Edd} = 4\pi G M_{\rm 2}m_{\rm p}c / \sigma_{\rm T}$
is the Eddington luminosity of the \Ms.

The dependence of $L_2$ on both gas density and relative velocity
means that the AGN luminosity of the \Ms\ may vary significantly
during each orbit and will also show a long-term evolution as the
orbit decays. Therefore, we expect the properties of the dAGN
population to change with redshift as MBH pairs evolve in time (see
also \citealt{LBB21a}). In this paper, we do not consider the instantaneous luminosity of dAGNs, instead, we use the evolution time weighted luminosity, which is calculated by summing the product of the instantaneous luminosity and the time step through the entire evolution, and then divide it by total evolution time.

To summarize, we have computed the full dynamical evolution (from
separations of $\approx 1$~kpc to coalescence) of 7988 MBH pairs
evolving in model post-merger galaxies derived from the TNG50-3
simulation. We also compute the \lisa\ SNR for those pairs that
successfully merge within a Hubble time. As described above, we also track estimates of the dAGN
luminosities of each pair while their separations are $\ga 1$~pc. In
the rest of the paper, we explore the connections between dAGNs and
MBH merger events. To do so, we define \zdagn\ as the redshift at
which a dAGN is observed and \zcoal\ as the redshift at which the two
MBHs that comprise the dAGN eventually coalesce (where \zcoal$ <$
\zdagn). The luminosity of a dAGN is defined as
$L_{\mathrm{bol}}=L_1+L_2$.

%%%%%%%%%%%%%%%%%%%%%%%%%%%%%%%%%%%%%%\
%  SECTION 3
%%%%%%%%%%%%%%%%%%%%%%%%%%%%%%%%%%%%%%\
\section{The Evolution of Dual Active Galactic
  Nuclei Luminosities and Separations}
\label{sec:AGN_dist_zla}
%In this section we present a statistical study of dAGNs that are precursors to MBH coalescences or \lisa\ sources. We present the separation, luminosity, and redshift distributions of dAGNs. These distributions provide a general view of all dAGNs in our study. We evolve the whole sample of TNG50-3 MBHBs in the inspiral phase before the coalescence for the four-year \lisa\ mission time and we evaluate the signal-to-noise ratio (SNR) of each coalescence using equation~(A4) of LBB22a. We describe the \lisa\ sensitivity and the calculation of GW SNR in the Appendix of LBB22a. In this study, we define a MBHB coalescence to be detectable by \lisa\ if its SNR is larger than 8, following the assumption in LBB22a.
In this section we provide an overview of the evolving dAGN population
found in our model suite.
%Out of the 2165 'merger' events identified
%from the TNG50-3 simulation, 45 have $M_{\rm gd}=0$ and 123 have both
%$M_{\rm gd}=0$ and $M_{\rm sb}=0$. These 168 systems are likely
%misidentified subhalos \citep{Ble2016, KBH2017, Katz2019} and are
%omitted from analysis as they will not host a dAGN. The dAGNs
%properties presented here result from $7988$ models calculated from
%the remaining $1997$ post-merger galaxies.
TNG50-3 predicts
a small number of galaxy mergers beyond a redshift of $3$ (Fig.~2 in
Paper I), so we limit our analysis of dAGNs to lower redshifts (in
particular, \zdagn$\leq 2$) where the population of dAGNs is largest.

\begin{figure}
 \includegraphics[width=0.5\textwidth]{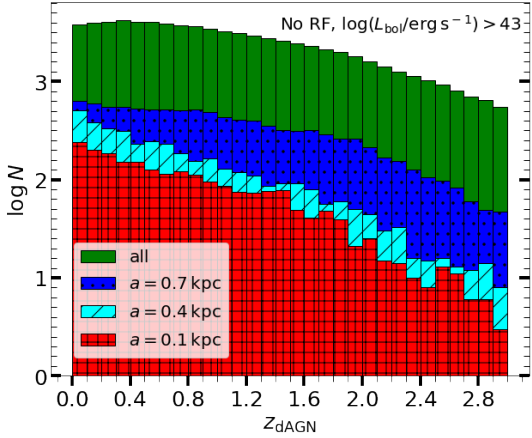}
 \caption{The redshift distribution of all dAGNs with
   total bolometric luminosity larger than $10^{43}$~erg~s$^{-1}$ and
   with separations of $0.7$, $0.4$, and $0.1$~kpc.}
\label{fig:z_dist}
\end{figure}

The minimum MBH
mass in the TNG50-3 simulation is $\approx 10^6$~M$_{\odot}$
\citep{TNG50_a}, which means that the lower-limit to the
dAGN luminosity distribution is $L_{\mathrm{bol}} \approx 10^{43}$~erg~s$^{-1}$
(i.e., the maximum allowed luminosity of a \Mp\ with this mass;
Eq.~\ref{eq:L1}). The solid green histogram in Figure~\ref{fig:z_dist}
shows how the total population of model dAGNs with $L_{\mathrm{bol}} >
10^{43}$~erg s$^{-1}$ evolves with \zdagn. Since the orbital evolution time of
a dAGN in the DF phase is frequently $>$Gyr (Paper I), one system can
appear in multiple redshift bins as long as its $L_{\mathrm{bol}} > 10^{43}$~erg~s$^{-1}$ in that redshift range.
We find that the largest number of dAGNs are present in the redshift range of
$0.4-0.5$, and the number drops as the redshift increases from $0.5$ to
$3$. The distribution of DF evolution times and galaxy gas fractions combine
to determine this redshift distribution. Paper I showed that the evolution time
for a \Ms\ to reach $a \sim 0.001$~kpc from $a \sim 1$~kpc ranges from $\sim
1$~Gyr (if the stellar bulge dominates) to $\sim 5-10$~Gyr (if the gas
disk dominates) depending on the orbital configuration of the
\Ms. Thus, only the systems containing dAGNs whose orbital evolution is
determined by stellar DF in the bulge and have relatively short
evolution time will evolve to pc-scales at high \zdagn. The dAGNs
whose orbital evolution is determined by the slower gaseous DF process
reaches this separation at lower redshifts.

%In this study we do not take into account of any kind of gas consumption and star formation in the host galaxies, which means that in the orbital decay from $\sim$ kpc to coalescence, the gas fraction of the host remains unchanged from the number inherited from TNG50-3 data file. For example, if a bright dAGN  ($L_{\rm bol}>10^{\rm 43} {\rm erg\, s^{\rm -1}}$) is at a redshift of $3$ with a separation of $0.7$ kpc, as the evolution goes on it will eventually become a brighter dAGN with a separation of $0.4$ kpc at lower redshift, unless it is stalled in the process. This is because as the \Ms\ evolves towards the galactic center, the gas density around the \Ms\ increases while that around the \Mp\ remains the same.

Fig.~\ref{fig:z_dist} also illustrates the redshift distribution of
dAGNs with separations of $0.7$, $0.4$ and $0.1$~kpc. All three
distributions peak at the smallest redshifts \zdagn\ $\approx
0$--$0.1$, but the peak is flattest and the total number of dAGNs
highest for $a=0.7$~kpc. This is because there are always more
dAGNs with large rather than with small separations. This happens because all
dAGNs with small separations were once dAGNs with large separations in
the past, but not all dAGNs with large separations evolve into dAGNs
with small separations. Taking everything into account, the shape of the redshift distribution of dAGNs is a natural result of DF dominated orbital evolution. This slow evolution means that \zdagn$\la 0.5$ is the
optimal region for observational searches for dAGNs. 

\begin{figure*}
  \centering
	\includegraphics[width=0.9\textwidth]{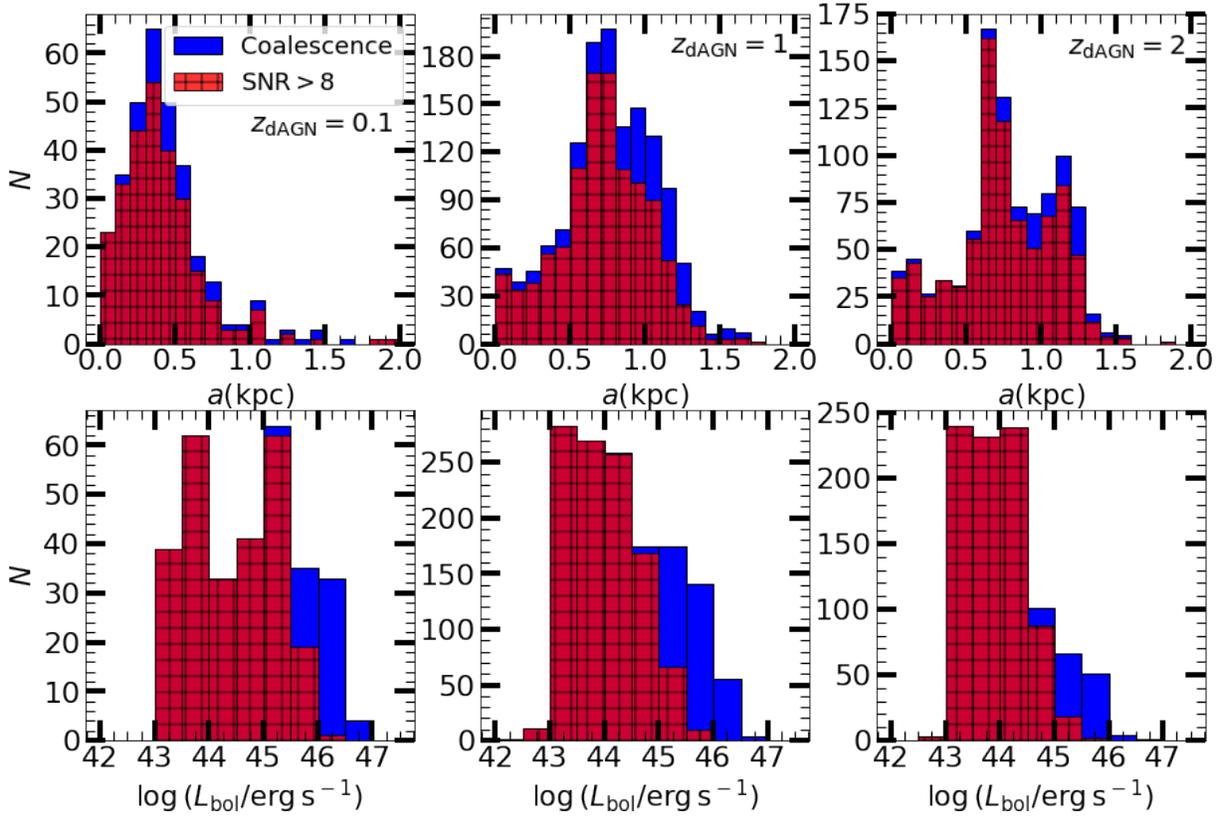}
\caption{The separation and luminosity distributions at \zdagn$=0.1$,
  $1$, and $2$ of dAGNs whose
  MBHs coalesce by \zcoal$=0$. The blue
  solid histograms show all dAGNs that eventually lead to coalescence by \zcoal$=0$, and the red hatched histograms show the
  systems in which the MBH merger has a \lisa\ SNR $>8$. }
\label{fig:al_dist}
\end{figure*}

Turning now to the 65\% of dAGNs that eventually lead to a
MBH merger, Figure~\ref{fig:al_dist} shows the separation and luminosity
distributions at \zdagn$=0.1$, $1$ and $2$ of all dAGN that will coalesce (solid blue histograms). The total number of dAGNs at \zdagn$=1$
and $2$ that eventually merge is larger than that at \zdagn$=0.1$
because MBH mergers can happen between \zdagn$=1$ and $0.1$ and these
do not appear in the leftmost distribution.
The top three panels of Fig.~\ref{fig:al_dist} show that most dAGNs at
\zdagn$=1$ and $2$ that eventually merge have separations in the range
of $0.5-1$~kpc. This is because the evolution time from $\sim
1$~kpc to coalescence is often in the range of $5-10$~Gyrs (Paper
I). The separation distribution at \zdagn$=2$ (the rightmost panel)
has a secondary peak at $a\sim 1.2$~kpc. This is because the time
between \zdagn$=2$ and $0$ ($\approx 10$~Gyrs) is long enough to allow
some MBH pairs with separations larger than $1$~kpc to merge within a Hubble time. The peak of the distribution at
\zdagn$=0.1$ is shifted towards $0.2-0.5$~kpc due to the relatively
short evolution time between \zdagn$=0.1$ and $0$ ($\sim 1$~Gyr).

The solid blue histograms in the bottom three panels of Fig.~\ref{fig:al_dist} show the
distribution of $L_{\mathrm{bol}}$ for all dAGNs at \zdagn$=0.1, 1,$ and $2$
whose MBHs merge before \zcoal$=0$. As expected, there is a sharp
cutoff at $L_{\mathrm{bol}} \approx 10^{43}$~erg~s$^{-1}$ due to the
minimum MBH mass in the TNG50-3 simulation. At all \zdagn\ the
majority of dAGNs that eventually merge have $L_{\mathrm{bol}} \approx
10^{43-45}$~erg~s$^{-1}$.

The hatched red histograms in Fig.~\ref{fig:al_dist} shows the
distributions of dAGNs that lead to \lisa\ SNR$> 8$. These potential
\lisa\ sources follow a similar trend in the separation distribution
at all three redshifts. However, at luminosities larger than $\approx
10^{45}$~erg~s$^{-1}$, the fraction of \lisa\ sources drops
significantly in the lower three panels. This is because the
bolometric luminosity is proportional to the binary mass, so more luminous precursor dAGNs have more massive MBHs and are less likely to be detected by
\lisa. %Thus, dAGNs that lead to MBH mergers and \lisa\ detections will
\section{The Coalescence Fraction of kpc-scale Dual Active Galactic Nuclei}
\label{sec:AGN_coal}
This section presents the coalescence fraction of dAGNs as a function
of redshift and observable properties of dAGN hosts. These fractions
provide a quantitative relation between the number of dAGNs and
merging MBHs, and can be used to estimate the cosmological
MBH binary coalescence rate and the number of \lisa\ detections from a sample of
detected dAGNs. 

\subsection{The Coalescence Fraction As a Function of Dual AGN Redshift}
\label{sec:coal_z}
\begin{figure*}
\centering
	\includegraphics[trim=5 80 5 85,clip,width=0.9\textwidth]{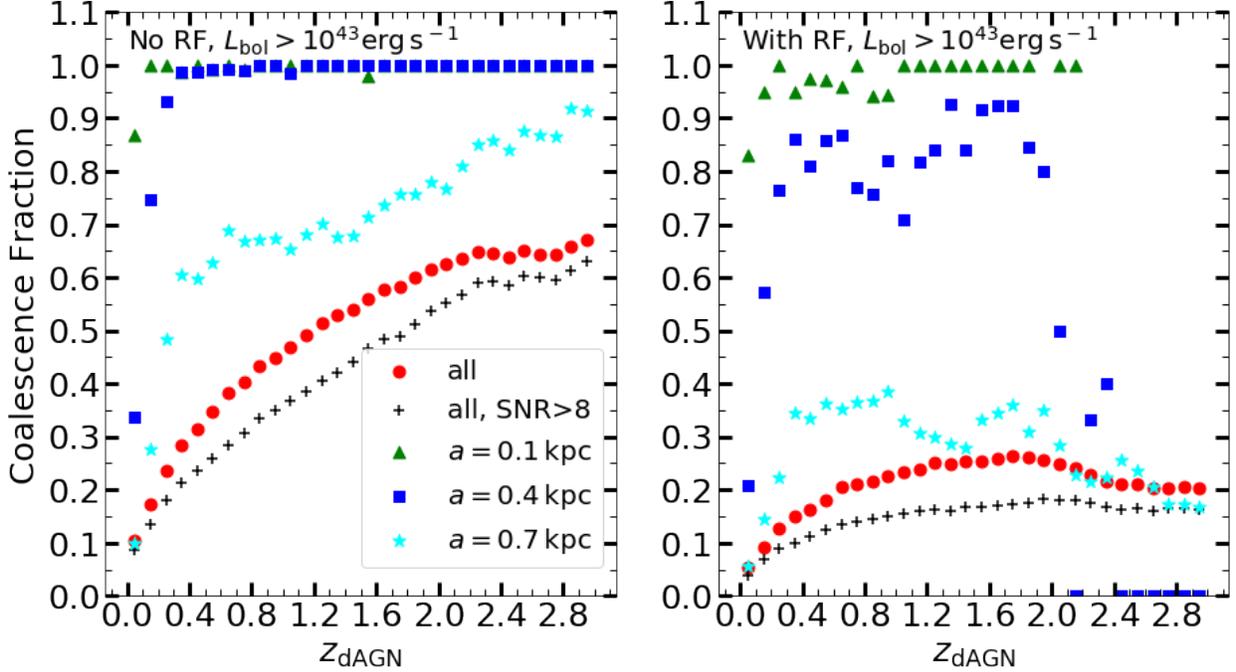}
\caption{The fraction of dAGNs observed at a given redshift whose MBHs
coalesce before \zcoal$=0$. The red circles show the fraction using
all 7988 dAGNs in the model suite, while the green triangles, blue
squares and cyan stars show how the fraction varies with dAGN
separation. The black '+' symbols indicate
the fraction of all dAGNs that lead to a \lisa\ SNR$>8$. The
right-hand panel shows how the fractions change when the effects of
radiation feedback (RF) are taken into account (\S~\ref{sec:RF_AGN_dist}).}
\label{fig:fraction}
\end{figure*}

In order to quantify the relationship between observable dAGNs and MBH
coalescence, we count the number of dAGNs with $L_{\mathrm{bol}} >
10^{43}$~erg~s$^{-1}$ with MBHs that coalesce before \zcoal$=0$ and
show the coalescence fraction as a function of \zdagn\ in the left
panel of Figure~\ref{fig:fraction}, grouped by dAGN separation. Almost all dAGNs with $a=0.1$~kpc
at \zdagn$>0.1$ eventually coalesce by \zcoal$=0$. The evolution time
of these systems is $10^{5-8}$~yrs (Paper I), which is sufficient for
a dAGN with a separation of $a=0.1$~kpc at \zdagn$=0.1$ to coalesce
(the corresponding cosmological time is $\sim 1.3$ Gyr). Thus, if a
dAGN is observed with $a\leq 0.1$~kpc at any redshift larger than $0.1$, then it has
nearly $100\%$ chance to coalescence in a Hubble time.

Similarly, all dAGNs with $a=0.4$~kpc at \zdagn$>0.4$ coalesce
before \zcoal$=0$, but the coalescence fraction between \zdagn$=0$ and
$0.1$ drops to $\sim 0.3$, showing that majority of these systems have
an evolution time longer than $\sim 1.3$~Gyrs. The coalescence
fraction of dAGNs with separations of $0.7$~kpc gradually increases as
redshift grows. This is as expected, since higher
redshift dAGNs that reach separations of $a=0.7$~kpc have a greater
probability to evolve to coalescence prior to
\zcoal$=0$. Indeed, the left panel of Fig.~\ref{fig:fraction} shows
that nearly $90\%$ of dAGNs with $a=0.7$ kpc at \zdagn$=3$ lead to a
MBH merger. However, this fraction drops to $50\%$ at
\zdagn$=0.3$. The coalescence fraction of dAGNs with separations of
$a=0.7$~kpc is in general lower than those with $a=0.1$ or $0.4$~kpc
because it is easier for the \Ms\ to evolve to coalescence from
$0.1$~kpc than from $0.7$~kpc.

The coalescence fraction of all dAGNs with $L_{\mathrm{bol}}>10^{43}$~erg~s$^{-1}$ is shown as the red circles in
Fig.~\ref{fig:fraction} and is $\sim 70\%$ at \zdagn$=3$ and $\sim
50\%$ at \zdagn$=1$. Thus, given a sample of \zdagn$=1$ dAGNs with
$L_{\mathrm{bol}}>10^{43}$~erg~s$^{-1}$ with separations of
$0.001-2$~kpc, we expect that half of them will coalesce before
\zcoal$=0$. In order to estimate the number of potential \lisa\ sources
using the number of observed dAGNs, we also count the number of dAGN
that coalesce before \zcoal$=0$ with \textit{LISA} SNR$>8$, and show
this fraction as black plus signs in the left panel of
Figure~\ref{fig:fraction}. As expected, the \lisa\ detectable fraction
is lower than the coalescence fraction at all redshifts, since
\lisa\ will not be as sensitive to the most massive MBH pairs.

%As shown by the red dotted line with circle markers, if some $L_{\rm bol}>10^{\rm 43} {\rm erg\, s^{\rm -1}}$ dAGNs with separations $< 2$ kpc are observed at $z\sim 0.7$, then $\sim 30\%$ of them may become \lisa\ sources. As expected, the \lisa\ detectable fraction is lower than the coalescence fraction at all redshifts and for all dAGN separations, since only some of the MBHBs can be detected with \lisa. 

%As shown in the figure, the difference between the \lisa\ detectable fraction and coalescence fraction for dAGNs with $0.7$ kpc separation is in general smaller than that for dAGNs with $0.4$ kpc separation. SNR is proportional to the chirp mass of a MBHB. Furthermore, at the same redshift, dAGNs which reach $0.4$ kpc separation have in general smaller binary mass than those reach $0.7$ kpc separation, which is a consequence of the hierarchical formation of galaxies. The difference between \lisa\ detectable fraction and coalescence fraction is small for dAGNs with large separation because these dAGNs are in general more massive.

\begin{figure*}[t]
\centering
	\includegraphics[trim=5 80 5 85,clip,width=0.9\textwidth]{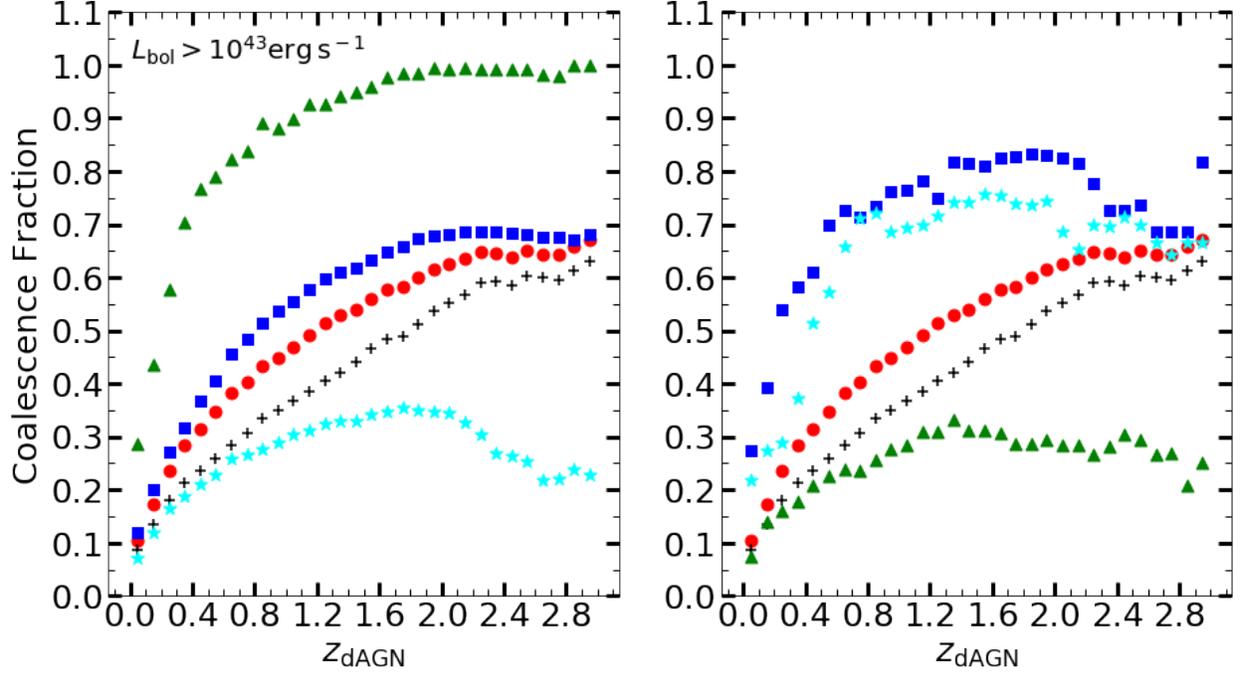}
\caption{The fraction of dAGNs at a given redshift that coalesce by
  \zcoal$=0$. Left: the coalescence fraction of dAGNs grouped by the bulge
  mass. Green triangles represent systems with $\log\,(M_{\rm
    sb}/\,M_{\rm \odot})=9$, blue squares represent systems with
  $\log\,(M_{\rm sb}/\,M_{\rm \odot})=10$, and cyan stars represent
  systems with $\log\,(M_{\rm sb}/\,M_{\rm \odot})=11$. Right: the
  coalescence fraction of dAGNs grouped by the mass ratio $q$. Green
  triangles represent systems with $\log\,q=-2$, blue squares
  represent systems with $\log\,q=-1$, and cyan stars represent
  systems with $\log\,q=-0.3$. In both panels, red circles illustrate
  the coalescence fraction of all dAGNs, and the black
  plus signs represent the fraction of all dAGNs that coalesce by
  \zcoal$=0$ and can be detected by \lisa.} 
\label{fig:fraction_mq}
\end{figure*}

Since the evolution time of the \Ms\ depends on the
properties of the post-merger galaxy and MBH pair
\citep[][Paper I]{LBB20a}, Figure~\ref{fig:fraction_mq} shows how
the coalescence fraction of dAGNs is impacted by the galaxy bulge mass (left
panel) and the MBH binary mass ratio (right panel). We find that the coalescence
fraction of dual AGNs is inversely proportional to
$M_{\mathrm{sb}}$, with the largest fractions in galaxies with bulge
masses of $M_{\mathrm{sb}} \approx 10^9$~M$_{\odot}$ and the lowest
with $M_{\mathrm{sb}} \approx 10^{11}$~M$_{\odot}$. This is a result
of the inverse relationship between $M_{\mathrm{sb}}$ and $f_g$
(\S~\ref{sec:galaxymodel}) so that galaxy models with less massive
bulges have higher gas fraction (and, thus, high gas densities and
sound speeds), where gaseous DF efficiently
decays the orbit of the \Ms\ at large separations (see also Paper I). When $M_{\mathrm{sb}}
> 10^{10}$~M$_{\odot}$, the gas
fraction falls to low enough values that gaseous DF becomes
inefficient, increasing the decay time of the \Ms\ before the stellar DF can takes over. 
The coalescence fractions of the entire dAGN population (red circles)
and the population that leads to \lisa\ SNRs$> 8$ (black plus signs)
closely follow the $M_{\mathrm{sb}} = 10^{10}$~M$_{\odot}$
results. This illustrates that galaxies with $M_{\mathrm{sb}} \approx
10^{10}$~M$_{\odot}$ dominate the sample of post-merger galaxies in
TNG50-3. The right panel of Fig.~\ref{fig:fraction_mq} shows that the
coalescence fraction of dAGNs with $q=0.1$ and $0.5$ have the highest
coalescence fractions because the DF forces are larger (and the
inspiral time shorter) for higher mass sMBHs \citep[][their Figure
  11]{LBB20a}. 

\subsection{The Cumulative Coalescence Fractions of Dual AGNs}
\label{sec:coal_para}
\begin{figure*}
\centering
\includegraphics[trim=5 90 5 90,clip,width=0.9\textwidth]{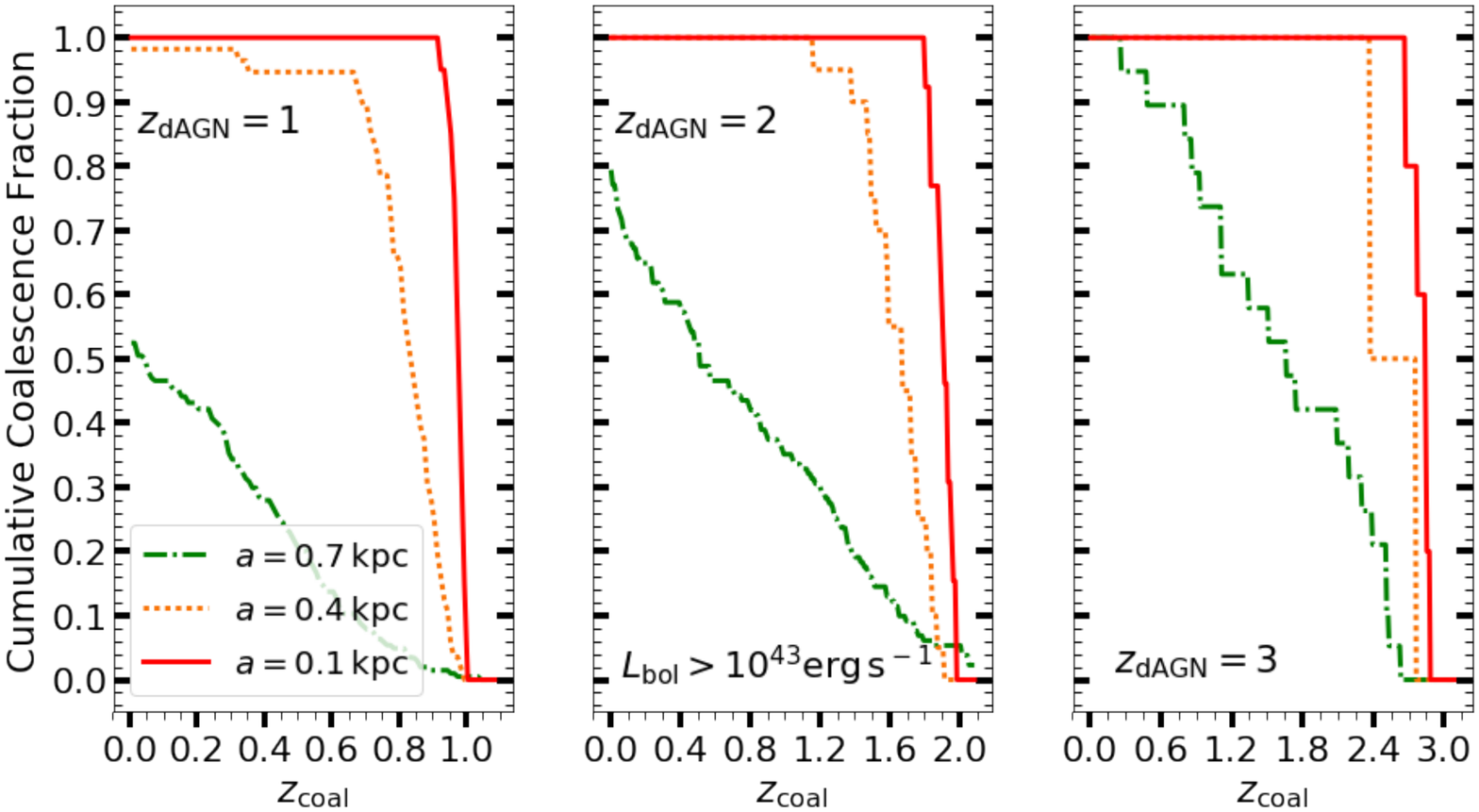}
\caption{The cumulative coalescence fraction (CCF) of dAGNs with
  $L_{\mathrm{bol}} > 10^{43}$~erg~s$^{-1}$ at \zdagn$=1$, $2$ and $3$.
  The dot-dashed green lines plot the CCFs of dAGNs with separations
  of $a=0.7$~kpc, while the dotted orange lines and
  solid red lines represent dAGNs with separations of $0.4$ and
  $0.1$~kpc, respectively.} 
\label{fig:CCF}
\end{figure*}

The results above focused on dAGNs that evolve to a MBH merger before
\zcoal$=0$, but it is also interesting to examine how these mergers
are distributed across redshift \zcoal.  Figure~\ref{fig:CCF} shows the
cumulative coalescence fractions (CCFs) of dAGNs with
$L_{\mathrm{bol}}  > 10^{43}$~erg~s$^{-1}$ at \zdagn$=1$, $2$ and
$3$.
\begin{figure*}
\centering
	\includegraphics[trim=5 50 5 30,clip,width=0.9\textwidth]{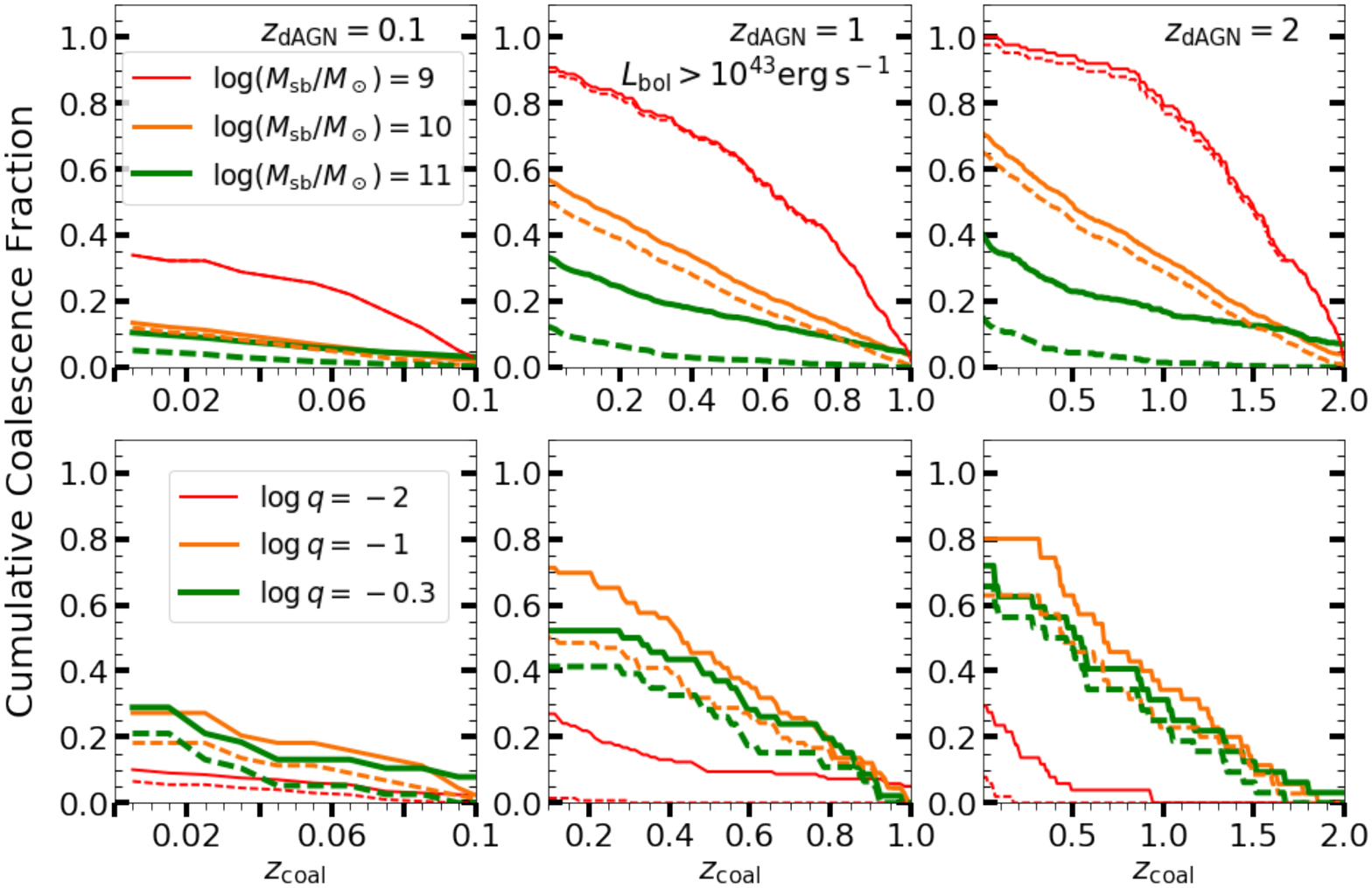}
\caption{The cumulative coalescence fraction (CCFs) of dAGNs at
  \zdagn$=0.1$, $1$ and $2$ with
  $L_{\rm bol}>10^{\rm 43} {\rm erg\, s^{\rm -1}}$ grouped by the
  stellar bulge mass (top row), and the MBH mass ratio (bottom row). The dashed lines show
  the CCFs of mergers that result in \lisa\ events with SNR$>8$.}
\label{fig:CCF_2}
\end{figure*}
The three lines in each panel shows the CCFs for dAGNs with different
separations at the starting \zdagn. Unsurprisingly, we see that dAGNs
that start with a separation of only $a=0.1$~kpc evolve to coalescence
efficiently, reaching a 100\% coalescence fraction by \zcoal$\approx
0.9$, $1.8$, $2.6$, if starting at \zdagn$=1$, $2$ and
$3$, respectively. Similarly, 100\% of the dAGNs with $a=0.4$~kpc in our sample that
are at \zdagn$=2$ or $3$ merge by \zcoal$\approx 1.2$
and $2.2$. Our calculations suggest that there may still
be a small fraction of the \zdagn$=1$, $a=0.4$~kpc population at low
$z$, but the majority of them would have merged by \zcoal$\approx
0.8$. In contrast, Fig.~\ref{fig:CCF} shows that dAGNs with separations of
$a=0.7$~kpc at \zdagn$=1$ or $2$ will not all have merged by
\zcoal$=0$ although the ones starting at \zdagn$=3$ would all have
merged by \zcoal$\approx 0.4$. Of course, the dAGNs that remain at low
redshifts will be at significantly smaller separations (e.g.,
Fig.~\ref{fig:al_dist}).

%All dAGNs above the luminosity threshold observed at $z_{\rm dAGN}=1- 3$ with separations smaller than or equal to $0.5$ kpc lead to coalescence by $z_{\rm coal}=0$. On the other hand, only $50\% - 80\%$ of $L_{\rm bol} > 10^{\rm 43} {\rm erg\, s^{\rm -1}}$ dAGNs observed with separations of $0.7$ kpc at $z=1-2$ lead to coalescence by $z=0$. Dual AGNs at higher redshifts have longer evolution time before $z=0$, and this is the reason for the higher cumulative coalescence fraction of all separations dual AGNs observed at higher redshifts. In summary, although the time when a dual AGN is observed is only a snapshot in its orbital evolution, duals that are identified at separations smaller than $0.5$ kpc and have high luminosities are generally promising indicators of coalescence. 

As in Fig.~\ref{fig:fraction_mq}, we show in Figure~\ref{fig:CCF_2}
how the CCFs vary with \Msb\ (upper panels) and $q$ (lower
panels). dAGNs of all separations are included in each curve, and the
dashed lines show CCFs of those mergers with \lisa\ SNRs$> 8$. We
replace the \zdagn$=3$ panels with one at \zdagn$=0.1$ to more closely
connect to observational searches for dAGNs.

%
%The top row of Figure~\ref{fig:CCF_2} illustrates the cumulative
%coalescence fraction of dAGNs with $L_{\rm bol}>10^{\rm 43} {\rm erg\,
%  s^{\rm -1}}$ observed at $z=0.1, 1,$ and $2$, and the effect of
%bulge mass on it. As shown in this figure, the CCF is in general
%higher for dAGNs observed at higher redshift. This is as expected
%since $\sim$ kpc scale dAGNs at higher redshift have more time to
%evolve before $z=0$, and hence are, more likely to coalesce compared
%to those dAGNs at lower redshift.
%In each panel of the top row of Figure~\ref{fig:CCF_2}, the CCFs of
%dAGNs in galaxies with $M_{\mathrm{sb}}=10^{\rm 9}$~M$_{\odot}$ bulge mass is always the
%highest, while that of dAGNs with $10^{11}$~M$_{\odot}$ bulge
%mass is always the lowest, consistent with the inverse relationship
%between $f_g$ and \Msb\ described 
%For a dAGN to have a bolometric luminosity larger than $10^{\rm 43} {\rm erg\, s^{\rm -1}}$, the gas density around the \Mp\ and \Ms\ needs to be larger than a certain threshold, which basically sets a threshold for the gas mass in the host galaxy. In TNG, lower mass galaxies have larger gas fraction on average, so the gas fraction of a dAGN in $10^{\rm 9} M_{\rm \odot}$ bulge is larger than that in a $10^{\rm 11} M_{\rm \odot}$ bulge, if these two dAGNs have similar bolometric luminosities. According to Figure~3 in LBB22a, the coalescence fraction of TNG50-3 MBHBs in high gas fraction systems is much larger than that in low gas fraction systems, which explains this trend in bulge mass. 
In the upper-row of the figure, we see the same inverse relationship
between \Msb\ and coalescence fractions as seen in
Fig.~\ref{fig:fraction_mq}. For example, $\sim 30$--$40$\% of
\zdagn$=0.1$ dAGNs in galaxies with $M_{\mathrm{sb}} =
10^{9}$~M$_{\odot}$ bulges will have coalesced by \zcoal$=0$. However,
if the dAGNs reside in post-merger galaxies with $M_{\mathrm{sb}} >
10^{10}$~M$_{\odot}$, then only $\sim 10$\% will merge in a Hubble
time. The fractions increase markedly when considering dual AGNs at higher redshifts, however the
fraction never rises above $40$\% for dAGNs in the most massive
bulges. In the bottom row of Fig.~\ref{fig:CCF_2}, we again see that
MBH pairs with larger values of $q$ will evolve faster and reach
coalescence at higher \zcoal\ than lower values of $q$. The implication
of this is that the most likely population of dAGNs persisting
to low redshift will be systems with $\log q \la -2$ and in galaxies
with $M_{\mathrm{sb}} \ga 10^{11}$~M$_{\odot}$. However, as shown by
the dashed lines, the strongest \lisa\ signals will originate from
MBH pairs evolving in lower mass bulges (see also Paper I).

\section{The Impact of Radiation Feedback Effects}
\label{sec:RF_AGN_dist}
The thermal pressure of the ionized bubble surrounding an accreting
MBH regulates its accretion rate \citep[e.g.][]{Ostriker1976,
  B1985, Ric2008, Park2011, Park2012} and suppresses its luminosity. The
magnitude of this effect depends on the motion of the MBH relative to
its gas environment, the gas density, and the temperature of
surrounding gas \citep{Park2013}. Meanwhile, the ionized bubble can
also reduce the DF force on MBHs moving in gas-rich hosts \citep{PB2017,
  G2020,T2020}, an effect known as ``negative DF''. In several previous papers we showed that these
radiation feedback (RF) effects may impact both the dAGN 
luminosities and the overall dynamical evolution of the \Ms\ \citep[][Paper
  I]{LBB20b,LBB21a}. In particular, Paper I showed that the negative
DF effect increases the orbital decay timescale of sMBHs in gas-rich
hosts, which could, in
principle, severely reduce the expected \lisa\ detection rates. In
this section, we explore how RF will impact the evolution of the dAGN
population. As the reduction in $L_{\mathrm{bol}}$ is relatively
modest (RF does not typically impact the luminosity of the pMBH;
\citealt{LBB21a}), we focus here on the RF effects on the orbital
decay of the \Ms. Details on how we implement RF in the calculation
of the DF force can
be found in the paper by \citet{LBB20b}. We note that the RF effects are only
computed during the DF phase of the MBH evolution, but, as seen Paper
I, this phase dominates the overall timescale of a MBH merger.

\begin{figure}
  \centering
  \includegraphics[trim=5 40 5 40,clip,width=0.45\textwidth]{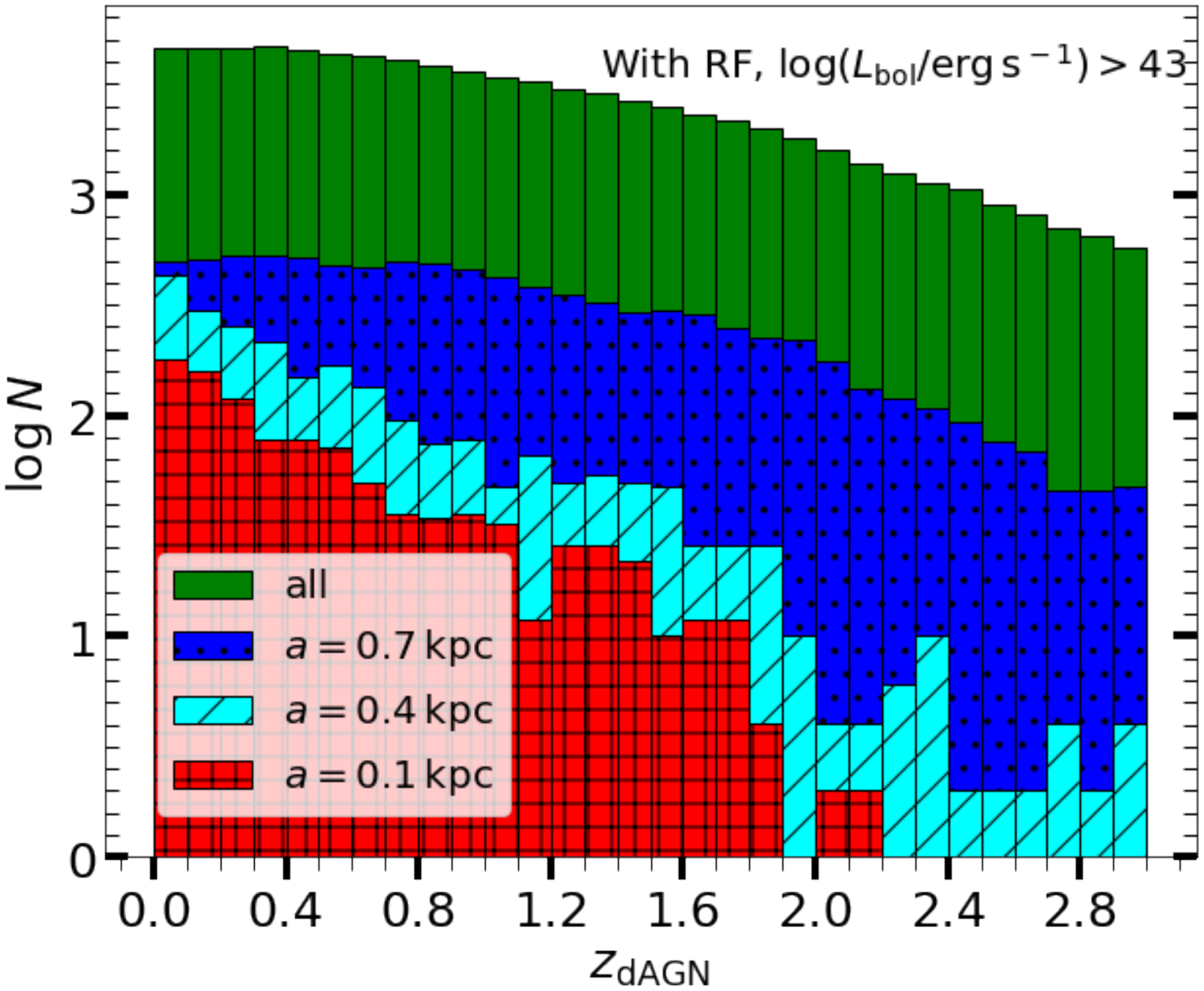}
\caption{As Figure~\ref{fig:z_dist}, but now showing the redshift
  distributions when including the effects of radiation feedback (RF). }
\label{fig:z_dist_RF}
\end{figure}

Figure~\ref{fig:z_dist_RF} shows the redshift distribution of all
dAGNs in our model suite with $L_{\rm bol} >10^{43}$~erg~s$^{-1}$ when
the RF effects are included in the calculations. This plot should be
compared to Fig.~\ref{fig:z_dist} which shows the results in the
absence of RF.
%Similar to
%Figure~\ref{fig:z_dist}, there are $2165\times 4$ of dAGNs in this
%figure since we associate each system with four orbital
%configurations. Furthermore, in this figure, each one of these
%aforementioned $2165\times 4$ dAGNs can appear in multiple bins of
%redshift as long as its $L_{\rm bol} > 10^{\rm 43} {\rm erg\, s^{\rm
%    -1}}$ in that redshift range.
%
The green histogram shows the redshift distribution of all dAGNs and
is similar to the one in Fig.~\ref{fig:z_dist}, except that there are
more dAGNs at $z \approx 0-0.4$ when RF effects are included. This is
because negative DF slows the orbital decay, which results in longer
evolution time and more dAGNs at low
redshifts. Figure~\ref{fig:z_dist_RF} also illustrates the redshift
distribution of dAGNs with separations of $a=0.7, 0.4$ and $0.1$~kpc
in the presence of RF. dAGNs separated by $a=0.7$~kpc still dominate
the population as in Fig.~\ref{fig:z_dist}. However, when RF effects
are included there are fewer
dAGNs with separations of $0.4$ and $a=0.1$~kpc, especially at large
redshifts. This is because negative DF increases the evolution time of
most systems with gas fractions larger than $0.1$
\citep{LBB20b}. Thus, it takes longer for most systems to reach smaller separations in the presence of RF, and thus, there are fewer dAGNs with separations $<0.4$ kpc at high redshifts when RF is taken into account.

\begin{figure*}[t]
\centering
	\includegraphics[trim=5 40 5 40,clip,width=0.9\textwidth]{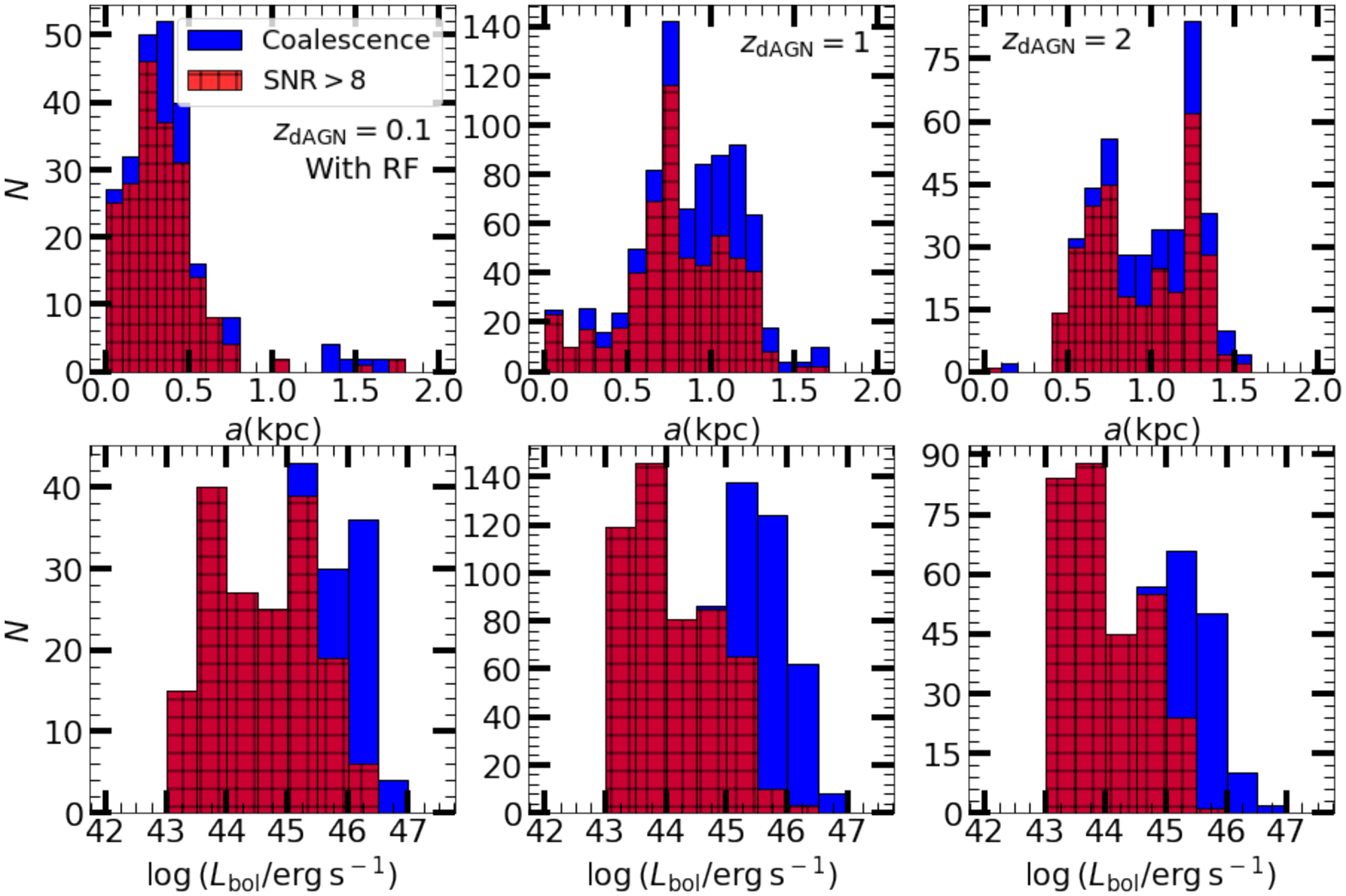}
\caption{As Figure~\ref{fig:al_dist}, but now showing the separation
  and luminosity distributions when RF effects are included.}
\label{fig:al_dist_RF}
\end{figure*}

The impact of RF on the separation and luminosity
distributions of all dAGNs that evolve to coalescence is shown in
Figure~\ref{fig:al_dist_RF} and should be compared to
Fig.~\ref{fig:al_dist}. Critically, the total number of systems that
reach coalescence decreases in all panels when RF effects are
included, with largest drop when \zdagn$=2$. This is a result of the
increase in evolution time caused by RF effects (see Paper I).
%The number of coalescence precursors
%peaks at $a=0.6-0.7$ kpc at $z=1$, which is smaller than that in the
%absence of RF ($a=0.7-0.8$ kpc). This is as expected, since the
%evolution time is increased by the RF, and a cosmological time of
%$\sim 7$ Gyrs from $z=1$ to 0 is only enough for dAGNs with smaller
%separations to coalesce.
%
%%%%%%%%%%%%%%%%%%%%%%%%%%%%%%%%%%%%%%\
%  FIGURE 9
%%%%%%%%%%%%%%%%%%%%%%%%%%%%%%%%%%%%%%\
%
%The separation distribution at $z=0.1$ shown in the left most panel of the top row of Figure~\ref{fig:al_dist_RF} is also slightly shifted towards smaller separations in the presence of RF, for the same reason. 
%  FIGURE 12
%%%%%%%%%%%%%%%%%%%%%%%%%%%%%%%%%%%%%%\
%\begin{figure*}[t]
%\centering
%	\includegraphics[trim=5 90 5 90,clip,width=0.9\textwidth]{fraction_tot_RF_2.pdf}
%\caption{Same as Figure~\ref{fig:fraction} but calculated in the presence of RF.}
%\label{fig:fraction_RF}
%\end{figure*}
%
The shape of the separation distributions are largely similar between
the two cases except for the one at \zdagn$=2$ which
remains bimodal in the presence of RF, but
the left peak at $a=0.6-0.7$~kpc is now lower than the right peak at
$a=1.2-1.3$~kpc, contrary to the outcome in the absence of RF. Dual
AGNs with a separation of $0.6-0.7$~kpc at \zdagn$=2$ tend to have
smaller binary masses and higher gas fractions compared to those dAGNs
with the same separation but observed at \zdagn$=1$. Although the
separation is the same, dAGNs at \zdagn$=2$ are more likely to stall
in the presence of RF than those at \zdagn$=1$. So the left peak at
$a=0.6-0.7$~kpc is visibly reduced. The right peak at $a=1.2-1.3$~kpc
is largely due to dAGNs whose orbital evolution is determined by
stellar DF in the bulge, which are also the ones that are least
affected by RF, so the right peak is only slightly reduced in the
presence of RF. 

The bottom row of Figure~\ref{fig:al_dist_RF} shows the bolometric
luminosity distribution of all dAGNs that eventually coalesce in the
presence of RF. Comparing this bottom row to the one in
Fig.~\ref{fig:al_dist} we find that most dAGNs that evolve to mergers
continue to have $L_{\mathrm{bol}}=10^{43-44}$~erg~s$^{-1}$ at all
\zdagn\ in the presence of RF, but the total number of such systems
has decreased significantly. However, the peak at
$L_{\mathrm{bol}}=10^{\rm 45-46}$~erg~s$^{-1}$ is barely affected by
RF. This is because $L_{\mathrm{bol}}$ is proportional to the MBH pair
mass, so the higher luminosity peak is composed of more massive dAGNs
whose orbital evolution is determined by stellar DF in the bulge and
is least affected by RF. Thus, the more luminous dAGNs are barely
affected by RF. The red-hatched histograms indicates systems that
evolve to MBH mergers with \lisa\ SNR$> 8$. As these are concentrated
in the lower mass, lower $L_{\mathrm{bol}}$, systems, RF effects
significantly decrease the number of these events (see also Paper I).

The impact of RF on the evolution of dAGNs is clearly seen in the
right panel of Fig.~\ref{fig:fraction} which shows the coalescence
fraction of dAGNs with $L_{\mathrm{bol}} > 10^{43}$~erg~s$^{-1}$ as a
function of \zdagn\ in the presence of RF. Almost all dAGNs with
separations of $a=0.1$~kpc at \zdagn$>0.1$ coalesce before \zcoal$=0$,
similar to the case in the absence of RF (the left panel of
Fig.~\ref{fig:fraction}). However, there are no $a=0.1$~kpc data
points at \zdagn$>1.8$, since negative DF increases the evolution
time, and none of our systems reaches $a=0.1$~kpc at these
redshifts. The square and star markers show the coalescence fraction
of dAGNs with 
$a=0.4$ and $0.7$~kpc, respectively. Comparing to the coalescence fraction in the absence
of RF as shown in the left panel of Fig.~\ref{fig:fraction}, at \zdagn$<1.8$, the coalescence fractions are in general
reduced by $10$--$20$\% for $a=0.4$~kpc dAGNs in the presence of RF,
while for $a=0.7$~kpc dAGNs the coalescence fraction is reduced by
$30$--$50$\%. At \zdagn$>1.8$, the coalescence fractions
are reduced significantly to $0.2$--$0.3$ no matter the dAGN
separation. This is because negative DF reduces the coalescence fraction.  

%Thus, the proportionality between the coalescence fraction
%and redshift in the absence of RF no longer exists in the presence of
%RF. 

Lastly, the red circles in the right panel of Fig.~\ref{fig:fraction}
represent the coalescence fraction of all dAGNs with $L_{\rm bol}>
10^{\rm 43} {\rm erg\, s^{\rm -1}} $ in the presence of RF. The
coalescence fraction at \zdagn$>0.7$ is nearly flat, and corresponds
to $0.2$--$0.25$. Thus, in the presence of RF, if dAGNs are observed
at \zdagn$>0.7$, then we expect $20$--$25$\% of them to coalesce by
\zcoal$=0$. The black plus signs in this plot illustrate the the
fraction of \lisa\ GW sources in the presence of RF. Similar to the
case in the absence of RF, the \lisa\ detectable fraction of dAGNs is
in general $\sim5$--$10$\% lower than the total coalescence fraction. 

\section{Discussion}
\label{sec:discuss}

\subsection{Predicting the MBH Merger Rate from Dual AGN
  Surveys}
\label{sec:method}
%%%%%%%%%%%%%%%%%%%%%%%%%%%%%%%%%%%%%%\
%  FIGURE 8
%%%%%%%%%%%%%%%%%%%%%%%%%%%%%%%%%%%%%%\
%\begin{figure*}[t]
%\centering
%            \includegraphics[width=0.9\textwidth ]{total_observation_curve.pdf}
%\caption{The observation rate of dAGNs with $L_{\rm bol} > 10^{\rm 43} {\rm erg\, s^{\rm -1}}$ as a function of observation redshift.}
%\label{fig:total_curve}
%\end{figure*}
%
The results presented above allow an estimate of the MBH merger rate,
and the subsequent \lisa\ detection rate,
to be derived from an observational survey of kpc-scale
dAGNs. Predictions can be calculated either in the presence or in the
absence of RF effects.
%Motivated by an increasing number of dAGNs detected with separations smaller few kpc, we consider how the results of this and similar theoretical studies can be used to estimate the cosmological coalescence rate and \lisa\ detection rate of MBHBs.  As shown in Figure~\ref{fig:z_dist}, most bright dAGNs with bolometric luminosity larger than $10^{\rm 43} {\rm erg\, s^{\rm -1}}$ exist at redshifts $0.1$ to $0.5$, and the number gradually drops beyond $z\approx 0.5 $. Furthermore, the redshift distribution of dual AGNs with separations smaller than $\sim 0.5$ kpc peaks sharply at $z\approx 0-0.1$. Figure~\ref{fig:fraction} shows that their coalescence fraction is larger at higher redshifts, reaching $\sim 70\%$ at $z=3$ and $\sim 50\%$ at $z=1$. These two pieces of information provide a way to connect, in a statistical sense, the number of EM detectable dAGNs and that of GW detectable MBHBs. 
%Using Figure~\ref{fig:fraction} observers can draw predictions for the rate of MBHBs emitting GWs in the \lisa\ band using the number of dAGN observations, which can serve as a roadmap for future searches for \lisa\ sources and bridge the gap between the EM and GW astronomy. 
Given a survey of dAGNs with $L_{\mathrm{bol}} >
10^{43}$~erg~s$^{-1}$ at \zdagn$\pm \Delta z$, the rate at which the MBHs in these systems
merge before \zcoal$=0$ is
\begin{equation}
  \label{eq:dAGN_rate}
  {dN_{\mathrm{coal}} \over dt}(z_{\mathrm{dAGN}}) =
  f_{\mathrm{coal}}\, n \,
  {4\pi c d^2_{L} \over (1+z_{\mathrm{dAGN}})^2},
\end{equation}
where $f_{\mathrm{coal}}$ is the coalescence
fraction at \zdagn\ from Fig.~\ref{fig:fraction}, $d_L$ is
the luminosity distance to \zdagn, and
$n$ is the observed dAGN comoving number density at \zdagn.
 % This quantity is calculated from the comoving number density
 % ($n_{\rm dAGN}(z)={\rm d}N(z)/\,{\rm d}V(z)$) of dAGNs with $L_{\rm
 % bol}>10^{\rm 43} {\rm erg\, s^{\rm -1}}$ using the formalism in
 % \citet{Haeh1994}. $d_{\rm L}(z)$ is the luminosity distance of the
 % coalescence. The co-moving number density of MBHBs as a function of
 % redshift is calculated from binning all $L_{\rm bol}>10^{\rm 43}
 % {\rm erg\, s^{\rm -1}}$ dAGNs observed by their redshifts (as shown
 % in Figure~\ref{fig:z_dist}), and dividing it by the observation
  % comoving volume at the corresponding redshift.
As an example, consider an all-sky survey that detected $400$
kpc-scale dAGNs with $L_{\mathrm{bol}} > 10^{43}$~erg~s$^{-1}$ at
\zdagn$=1\pm0.01$. Then
\begin{enumerate}
  \item the dAGN comoving number density is $n=400/4\pi d_L^2c
    dt$. The time interval $dt$ is the cosmological time across
    \zdagn$=1\pm0.01$ which is $0.081$~Gyr in our adopted cosmology (see \S~1).
  \item According to the left panel of Fig.~\ref{fig:fraction} the
    coalescence rate for dAGNs at \zdagn$=1$ in the absence of RF
    effects is $f_{\mathrm{coal}}\approx 0.45$. The fraction of dAGNs
    that lead to a merger with a \lisa\ SNR$>8$ is $\approx 0.35$.
  \item The coalescence rate of these dAGNs at \zdagn$=1$ is found
    from Eq.~\ref{eq:dAGN_rate}: $dN_{\mathrm{coal}}/dt \sim 6\times
    10^{-7}$~yr$^{-1}$. The rate of mergers with a \lisa\ SNR$>8$ is
    $\sim 5\times 10^{-7}$~yr$^{-1}$.
\end{enumerate}
%
%$f_{\rm coal}(z=1) n_{\rm dAGN}(z=1)\times \frac{4\pi c d^{\rm
%2}_{L}(z=1)}{(1+1)^{\rm 2}}$, where $f_{\rm coal}(z=1) \approx 0.45$
%according to the left panel of Figure~\ref{fig:fraction}, and hence
%the coalescence rate at $z=1$ is $\sim 6\times 10^{\rm -7}\, {\rm
%yr^{\rm -1}}$ in this example. The total coalescence rate can be
%obtained by integrating ${\rm d} N_{\rm coal} (z) /{\rm d} t$ over
%all redshifts. In Chapter 5, we estimate the MBHB coalescence rate to
%be $\sim 0.45$ ${\rm yr^{-1}}$ and the $\textit{LISA}$ detection rate
%to be $\sim 0.34$ ${\rm yr^{-1}}$ using similar method. A prediction
%of coalescence rate in the presence of RF can be estimated in a
%similar way using the coalescence fraction in the right panel of
%Figure~\ref{fig:fraction}.
To include the effects of RF, the calculation can be repeated using
the values of $f_{\mathrm{coal}}$ from the right-hand panel of
Fig.~\ref{fig:fraction}. In the example above, RF reduces the
predicted merger rates to $3\times 10^{-7}$~yr$^{-1}$ (all systems) and $2\times
10^{-7}$~yr$^{-1}$ (those with \lisa\ SNR$>8$).

Eq.~\ref{eq:dAGN_rate} gives the expected rate of MBH mergers from a
population of dAGNs at a particular \zdagn. If this estimate can be
made at multiple \zdagn\, the results can be integrated to yield the
total MBH merger rate. In Paper I we found that the overall MBH
coalescence rate is $\sim 0.45$~yr$^{-1}$ in the absence of RF
effects, while the rate of sources with a \lisa\ SNR$>8$ is $\sim
0.34$~yr$^{-1}$.  Comparing the merger rates derived from dAGN
observations and Eq.~\ref{eq:dAGN_rate} with future
\lisa\ measurements will provide important constraints on the
efficiency of DF forces and RF effects in the orbital evolution of MBH pairs.

\subsection{Comparison to Results from Literature}
\label{sub:comparison}
%Now we compare our results to that of \citet{Marta2021}.
In the paper by \citet{Marta2021}, dAGNs in the cosmological
simulation $\textit{Horizon-AGN}$ are identified and related to the
corresponding MBH binary mergers from the same simulation
\citep{Marta2020}, and the numerical relation between the dAGNs and
MBHB mergers is studied. The right column of their Figure~11
illustrates the CCFs of dAGNs with $5\sim 10$ kpc
separations. According to their results, $30-60\%$ of these dAGNs
observed at $z_{\rm dAGN}\sim1- 3$ coalesce before \zcoal$=0$,
increasing with decreasing $z_{\rm dAGN}$. This is due to the long
evolution time of dAGNs with small mass ratios at high
redshifts. These high redshift dAGNs with small mass ratios are formed
in \textit{Horizon-AGN} simulation because the criterion for MBH
formation is based only on gas properties. Thus, at high redshifts,
when the gas reservoir is rich, some MBHs form in gas clouds that are too small to be identified as galaxies ('intergalactic' MBHs). These small 'intergalactic' MBHs can be captured by galaxies later on and shine as dAGNs if some stochastic accretion occurs \citep{Marta2021}. According to the left column of Figure~11 of \citet{Marta2021}, dAGNs with one 'intergalactic' MBH have small mass ratios and occur frequently at high redshifts ($20\sim 30\%$ of all dAGNs at $z=2, 3$). The evolution time of these systems is long and the resulting coalescence fractions are low due to the small mass ratios, which leads to low coalescence fractions of dAGNs at high redshifts as shown in the right column of their Figure~11.

%This is because star formation rates in host galaxies at larger redshifts are in general higher, resulting in smaller gas reservoirs and longer dynamical friction time \citep{Marta2020}. Overall, the results of \citet{Marta2021} indicate dAGNs that are identified at small separations are generally indicators of effective mergers, which is in agreement with our results.

In comparison with Figure~\ref{fig:CCF} of this paper, the CCFs in \citet{Marta2021} are in general lower at all $z_{\rm dAGN}$. This is because the initial dAGN separations considered in \citet[][$30-50$\,kpc]{Marta2021} are larger than that considered in this work ($\sim $ kpc), so the evolution time is longer and CCFs are lower in their case. As shown in Figure~\ref{fig:CCF}, the CCFs increase with increasing $z_{\rm dAGN}$, contrary to the results of \citet{Marta2021}. This is because we do not include those  systems comprising one 'intergalactic' MBH in the analysis of this work (see \S~3 of paper I). Overall, the results of \citet{Marta2021} indicate dAGNs that are identified at small separations are generally indicators of effective mergers, which is in agreement with our results.

%we do not include any stellar formation in host galaxies during the evolution. If taking the stellar formation and corresponding gas consumption into account, the gas fractions of galaxies at larger $z_{\rm dAGN}$ would be lower, the evolution time would be longer, and the resulting CCFs would be lower than that shown in Figure~\ref{fig:CCF}. If taking the stellar formation into account, we expect the CCFs of dAGNs at large $z_{\rm dAGN}$ to be affected the most. Other impact of this assumption is discussed in the next section.

\subsection{Impact of Simplifying Assumptions}
\label{sub:assumptions}

The advantage of our semi-analytic model is its ability to provide calculations over a wide range of galaxy and MBH orbital properties at the cost of making some simplifying assumptions. The potential impact
of our assumptions on the dynamical evolution of MBH pairs is
discussed by \citet[][Paper I]{LBB20a,LBB20b}. In this section, we
consider the possible effects of these assumptions on the dAGN
properties and their connection to MBH mergers.

We assume that the \Mp\ is fixed at the center of the host galaxy. If
the motion of the \Mp\ and its corresponding Bondi-Hoyle-Lyttleton
accretion rate are accounted for in the calculations, the resulting
bolometric luminosity would be lower and the evolution time would be
shorter. In this case, there would be fewer dAGNs with
$L_{\mathrm{bol}} > 10^{43}$~erg~s$^{-1}$ and more MBH mergers at
higher redshift, resulting in a larger coalescence fraction of
dAGNs. Consequently, including the motion of the \Mp\ would increase
the \lisa\ detection rate. This effect would be strongest in
comparable mass MBH pairs and weaker in those with small $q$. It would
be manifested as an increase in the number of high redshift
\lisa\ detections, since MBH pairs at high redshifts tend to have
larger mass ratios.
%However, the impact on the EM detectability of precursor dAGN can not be determined without running more detailed simulations.

The orbit of the sMBH is assumed to always reside in the midplane
of the model remnant galaxy. If an inclined orbit takes the
\Ms\ outside of the galactic gas disk, the evolution time would
increase and $L_{\mathrm{bol}}$ would decrease. We also assume that
the sMBHs do not grow in mass during their orbital evolution from kpc
scales toward coalescence. Had they been able to do so, the increase
in the total mass of the binary would render the inspiral time shorter
and their bolometric luminosity higher. 

As mentioned in \S~\ref{sec:RF_AGN_dist}, we considered the effect
of RF only during the DF phase. Some studies have shown that RF can
also affect the orbital evolution of sMBHs in circumbinary
disks if there is no gap formed. Especially when sMBHs accrete at high rates, the radiation leads to strong
winds pushing against the gas disk, blowing the gas
away from the binary which stalls the binary hardening in the circumbinary disks \citep{RF1pc, William22}. On the contrary, when there is a gap in the disk, RF does not affect the evolution time of sMBHs \citep{RF1pc}. Our model assumes the gap-opening
regime (Paper I), so taking into account RF in circumbinary disks should not significantly affect the dAGN coalescence fraction and the $\textit{LISA}$ detection fraction predicted in this work.

Triplets of MBHs are likely to form at high redshifts, when the merger rate of galaxies is high. MBHs in these triplets may undergo the Kozai-Lidov oscillations which may increase the eccentricity of the central MBHBs  \citep{Kozai1962} and increase the dAGN coalescence rate. Besides the Kozai-Lidov oscillations, the chaotic three-body interactions can also boost the coalescence rate \citep{Bla2002,Hoffman2007, Amaro2010, Kul2012, Bonetti2016, Ryu2018}. If triplets and the three-body interactions were included in our model, the dAGN coalescence fractions and the $\textit{LISA}$ detection fraction would be higher.

We do not take into account gas consumption and star formation in the
merger remnant galaxies. This means that during the orbital evolution
from $\sim$~kpc to coalescence, the gas fraction of a host remains at
the same value inherited from TNG50-3 data file. This assumption
potentially increases the number of dAGNs with
$L_{\mathrm{bol}}>10^{43}$~erg~s$^{-1}$, since in reality at least a
fraction of the gas reservoir turns into stars. This reduces the gas
fraction and increases the stellar density in the host galaxies. In the
presence of RF, higher stellar densities result in more efficient
orbital evolution of MBH pairs and more MBH mergers. Thus, in the
presence of RF the coalescence fraction of dAGNs would be higher than
determined here. In the absence of RF, the effect of this assumption
can not be easily predicted due to the complicated interplay of DF and
the galactic parameters \citep{LBB20a}. The host galaxies with high gas
fraction are affected most by this assumption. 

%%%%%%%%%%%%%%%%%%%%%%%%%%%%%%%%%%%%%%\
%  SECTION 7
%%%%%%%%%%%%%%%%%%%%%%%%%%%%%%%%%%%%%%\
\section{Conclusions}
\label{sec:concl}
dAGNs are a product of galaxy mergers and trace
the population of future MBH coalescences. In this paper, we combined the
MBH dynamical evolution calculations from Paper I with estimates of
AGN luminosity to explore how the luminosities and
separations of dAGNs change as the MBH pair evolves in its
host galaxy. In addition, we were able to calculate the fraction of the
dAGN population at a redshift \zdagn\ that lead to a MBH merger at
redshift \zcoal, including determining the fraction that lead to a
\lisa\ SNR$> 8$.

We find that, in the absence of RF effects, the dAGN population in our model, with total
bolometric luminosity $L_{\mathrm{bol}} > 10^{43}$~erg~s$^{-1}$, peaks at
\zdagn$\approx 0.4$ and is dominated by systems with separations $a \ga 0.7$~kpc (Fig.~\ref{fig:z_dist}). However, a majority of
these dAGNs will not lead to MBH mergers before \zcoal$=0$
(Fig.~\ref{fig:fraction}). In fact, the majority of low-\zdagn\ dAGNs that
are precursors to MBH mergers are separated by $\la 0.5$~kpc
(Fig.~\ref{fig:al_dist}). This is a result of the orbital decay times
of the \Ms -- there is simply not enough time for most $a\sim 0.7$~kpc
dAGNs at \zdagn$\approx 0.4$ to evolve through to coalescence before
\zcoal$=0$. Therefore, a closer connection between dAGNs and MBH
mergers can be most easily obtained by detecting dAGNs at \zdagn$\ga
1$, where the coalescence fraction exceeds $0.5$. The separation of
dAGNs at \zdagn$\ga1$ is $\sim 0.7$--$1$~kpc and the total
$L_{\mathrm{bol}}\approx 10^{43-46}$~erg~s$^{-1}$.

The orbital evolution of the \Ms\ depends on the properties of the
post-merger galaxy and MBH pair \citep[e.g.,][Paper I]{LBB20a}, so
these conditions also impact the coalescence fractions of dAGNs. We
find that dAGNs in post-merger galaxies with bulge masses
$M_{\mathrm{sb}} \la 10^{10}$~M$_{\odot}$ and with MBH mass ratios of
$q \approx 0.5$ have the highest coalescence fractions. The fractions increase with $z$, observational searches for dAGNs that
lead to MBH mergers may consider prioritizing galaxies with less
massive stellar bulges.

In Paper I we found the radiation feedback effects can significantly
increase the evolution timescales for MBH pairs, in particular in high
gas fraction galaxies. This phenomenon (``negative DF'') leads to a
drop in the expected MBH coalescence rate and is seen in the dAGN
properties when we include this effect (Fig.~\ref{fig:fraction}
and~\ref{fig:al_dist_RF}). A signature of RF effects is in a
larger than expected number of dAGNs with wide separations, as this
would indicate slow orbital decay predicted by negative DF.

The coalescence fractions shown in Fig.~\ref{fig:fraction} can be
combined with the results of dAGN surveys to predict both the MBH
merger rate and the rate of \lisa\ signals with SNR$> 8$. We provide a
recipe in \S~\ref{sec:method} that can be followed to calculate
these rates (either with or without RF effects) from observations of
dAGNs. Comparing these predicted rates to \lisa\ measurements can be used to test our understanding of DF, including the importance of RF effects. The results like the ones presented here, in combination with the next-generation of dAGN
surveys, will thus be crucial in testing the physical models of MBH
evolution at sub-kpc scales.

\acknowledgments

T.B. acknowledges the support by the National Aeronautics and Space Administration (NASA) under award No. 80NSSC19K0319 and by the National Science Foundation (NSF) under award No. 1908042.

\end{document}